\documentclass[twocolumn]{IEEEtran} 
\usepackage{amsmath,amssymb,amsfonts}
\usepackage[final]{graphicx}
\usepackage{psfrag}
\usepackage[tight]{subfigure}
\usepackage[numbers,sort&compress]{natbib}
\usepackage{multicol}
\usepackage{mathrsfs} 
\usepackage{epstopdf}
\usepackage{multirow}
\usepackage{amsthm,mathtools}
\usepackage{float}
\usepackage{booktabs}  
\usepackage{multirow} 
\usepackage{relsize}
\usepackage{bm}
\usepackage{hyperref}
\usepackage{tabularx}

\usepackage[nolist,nohyperlinks]{acronym}
\usepackage{svg}
\usepackage{xparse}
\NewDocumentCommand{\evalat}{sO{\big}mm}{%
  \IfBooleanTF{#1}
   {\mleft. #3 \mright|_{#4}}
   {#3#2|_{#4}}%
}
\newcommand{\norm}[1]{\left\lVert#1\right\rVert}

\begin{acronym}
\acro{PLS}{physical layer security}
\acro{OP}{outage probability}
\acro{FRIS}{\emph{Fluid RIS}}
\acro{SNR}{signal-to-noise ratio}
\acrodefplural{SNR}[SNRs]{signal-to-noise ratios}
\acro{AWGN}{additive white Gaussian noise}
\acro{CSI}{channel state information}
\acro{PDF}{probability density function}
\acrodefplural{PDF}[PDFs]{probability density functions}
\acro{CDF}{cumulative distribution function}
\acro{PDF}{probability density function}
\acrodefplural{CDF}[CDFs]{cumulative distribution functions}
\acro{LRS}{large reflecting surface}
\acro{BS}{base station}
\acro{RF}{radio frequency}
\acro{RV}{random variable}
\acro{FN}{folded normal}
\acro{TDD}{time-division duplexing}
\acro{LOS}{line-of-sight}
\acro{DL}{downlink}
\acro{MRT}{maximal ratio transmission}
\acro{MRC}{maximal ratio combining}
\acro{MC}{Monte Carlo}
\acro{RV}{random variable}
\acro{CDF}{cumulative density function}
\acro{FAS}{fluid antenna system}
\acro{MIMO}{multipe-input multiple-output}
\acro{BS}{base station}
\acro{CSI}{channel state information}
\end{acronym}

\newtheorem{proposition}{Proposition}

\newcommand{\diag}{\mathop{\mathrm{diag}}}

\usepackage[numbers,sort&compress]{natbib}

\makeatletter
\def\blfootnote{\xdef\@thefnmark{}\@footnotetext}
\makeatother

\usepackage{float}
\usepackage{stfloats}
\usepackage{textcomp}
\usepackage{algpseudocode}
\usepackage{algorithm}

\begin{document}
\title{\huge{Exploring Spatial Flexibility and Phase Design in Fluid Reconfigurable Intelligent Surfaces: A Physical Layer Security Perspective}}
\author{J.~D.~Vega-S\'anchez, \textit{Senior Member, IEEE}, V. H. Garz\'on Pacheco, N. V. Orozco Garz\'on, \textit{Senior, IEEE}, D. A. Riofrío Almeida, \textit{Member, IEEE}, and D.~P.~Moya~Osorio, \textit{Senior Member, IEEE} }

\maketitle

\blfootnote{\noindent Manuscript received MONTH xx, YEAR; revised XXX. The review of this paper was coordinated by XXXX.  \\
(\textit{Corresponding author: Jos\'e~David~Vega-S\'anchez})}

\blfootnote{\noindent 
J.~D.~Vega-S\'anchez is with the Colegio de Ciencias e Ingenier\'ias  ``El Polit\'ecnico", Universidad San Francisco de Quito (USFQ), Diego de Robles S/N, Quito (Ecuador) 170157.
}

\blfootnote{\noindent
V. H. Garzón Pacheco and N.~V.~Orozco~Garzón are with the Faculty of Engineering and Applied Sciences (FICA), Telecommunications Engineering, Universidad de Las Am\'ericas
(UDLA), Quito 170124, Ecuador. 
}

\blfootnote{\noindent
D.~P.~Moya~Osorio is with Communication Systems Division, Department of Electrical Engineering, Linköping University, 581 83 Linköping, Sweden.
}

\vspace{-12.5mm}
\begin{abstract}
This work examines the secrecy outage probability (SOP) in Fluid Reconfigurable Intelligent Surfaces (FRIS) and contrasts their performance against two alternative RIS architectures: a traditional planar RIS and a compact RIS layout. To characterize the end-to-end FRIS channel, a maximum likelihood estimation (MLE) approach is introduced, while a Q-learning algorithm is employed to adaptively select the spatial positions of FRIS elements. Numerical evaluations show that optimizing element placement in FRIS significantly improves SOP compared to conventional RIS without phase adaptation. However, these improvements become less evident once the conventional RIS implements optimized beamforming (BF) and phase-shift (PS) controlling. In addition, FRIS maintains a clear advantage over compact RIS designs with optimized BF and PS, mainly due to its lower spatial correlation. Results further indicate that reducing the inter-element distance negatively impacts SOP, highlighting the importance of spatial diversity.
\end{abstract}


\vspace{-2.5mm}
\section{Introduction}
Reconfigurable Intelligent Surfaces (RISs) have emerged as a promising approach for achieving cost-effective and energy-efficient control of the radio environment. By leveraging nearly passive reflecting elements with tunable phase shifts, RISs can reshape wireless channels to allow gains in coverage and performance, especially in scenarios where direct links are severely weakened or blocked \cite{Abdelhamid}. Despite their potential, conventional RIS implementations are limited by spatial correlation, discrete phase constraints, restricted degrees-of-freedom (DoF), and control overhead. These issues may limit their performance in practice. Despite their potential, conventional RIS implementations are limited by spatial correlation, discrete phase constraints, and restricted degrees-of-freedom (DoF), and control overhead associated with configuring a large number of reflecting elements. These issues may limit their performance in practice.

Considering the potential advantages of RIS, the concept of fluid reconfigurable intelligent surfaces (FRISs) has recently been introduced \cite{Junjie}. Unlike fixed-layout RISs, FRIS allows the reflecting elements to physically relocate within predefined subregions, adding spatial position optimization as a new DoF. This flexibility enables FRIS to significantly improve achievable rates, sensing performance, and interference management compared to conventional RISs. 
Furthermore, extensions such as Fluid Integrated Reflecting and Emitting Surfaces (FIRES) broaden the range of applications by combining reflection and emission functionalities \cite{Rostami}. 
In parallel, the study of FRIS within the context of physical layer security (PLS) has gained increasing attention, motivated by the inherent advantages of movable antennas in enhancing secrecy performance. For instance, the work in~\cite{MovablePLS1} analyzed the fundamental principles and achievable secrecy rates of movable antennas under different channel conditions. In~\cite{MovablePLS2}, the authors investigated the secrecy outage probability (SOP) and demonstrated that spatial adjustment effectively mitigates eavesdropping. The approach in~\cite{MovablePLS3STARTS} provided a statistical analysis showing the overall secrecy gains achievable through movement diversity. Such works highlight that, while movable or fluid antennas can exploit spatial randomness to enhance PLS, their benefits come at the cost of increased control and channel estimation overhead.
 In parallel, the study of FRIS within the context of physical layer security (PLS) has gained increasing attention, motivated by the inherent advantages of movable antennas in enhancing secrecy performance \cite{MovablePLS1, MovablePLS2, MovablePLS3STARTS}. By dynamically adjusting element positions, FRIS can exploit spatial randomness to reduce the information leakage towards eavesdroppers, thereby lowering the secrecy outage probability (SOP). Such advantages become particularly relevant in scenarios where traditional RISs face limitations due to strong spatial correlation.

Motivated by these developments, this paper investigates the SOP performance of FRIS-assisted systems under full optimization of spatial positions, beamforming (BF), and phase shifts (PS). In contrast to previous works that rely on statistical approximations or unsupervised clustering-based methods for channel modeling \cite{Abdelhamid}, we adopt a closed-form maximum likelihood estimation (MLE) framework to characterize the end-to-end channel. Moreover, we employ a Q-learning algorithm to efficiently optimize the element positions of FRIS, while BF and PS are designed through a distributed optimization scheme. Our results reveal that although FRIS provides significant SOP gains compared to conventional RIS without phase optimization, such benefits diminish when conventional RIS employs fully optimized BF and PS. Nevertheless, FRIS consistently outperforms compact RIS architectures due to its reduced spatial correlation, and we further demonstrate that SOP degrades as the inter-element spacing decreases.

\vspace*{2mm}
\noindent\textit{Notation and terminology:} Uppercase and lowercase bold letters denote matrices and vectors, respectively;$f_{(\cdot)}(\cdot)$ is a probability density function (PDF); $F_{(\cdot)}(\cdot)$ is a cumulative density function (CDF); $\mathcal{C}\mathcal{N}(\cdot ,\cdot )$ is the circularly symmetric complex Gaussian distribution; $J_0(\cdot)$ is  the zeroth-order Bessel function of the
first kind; $\left ( \cdot \right )^{\rm H}$ is the Hermitian transpose;  $\left ( \cdot \right )^{\rm T}$ is the transpose.

\vspace{-4mm}
\section{System and Channel Models}
\begin{figure}[t]
\centering
\psfrag{A}[Bc][Bc][0.7]{{Alice}}
\psfrag{B}[Bc][Bc][0.7]{{Controller}}
\psfrag{C}[Bc][Bc][0.7][-15]{{FRIS}}
\psfrag{D}[Bc][Bc][0.7][-10]{$\mathbf{G}$}
\psfrag{E}[Bc][Bc][0.7][-15]{$\Psi$}
\psfrag{I}[Bc][Bc][0.7][72]{${\bf{h}}_{2,\mathrm{B}}$}
\psfrag{J}[Bc][Bc][0.7][92]{${\bf{h}}_{2,\mathrm{E}}$}
\psfrag{F}[Bc][Bc][0.7]{{Bob}}
\psfrag{M}[Bc][Bc][0.7]{{Eve}}
\psfrag{G}[Bc][Bc][0.6]{Preset~Position}
\psfrag{H}[Bc][Bc][0.6]{{Selected~Position}}
\psfrag{K}[Bc][Bc][0.7]{Main channel} 
\psfrag{L}[Bc][Bc][0.7]{Wiretap channel} 
\includegraphics[scale=0.24]{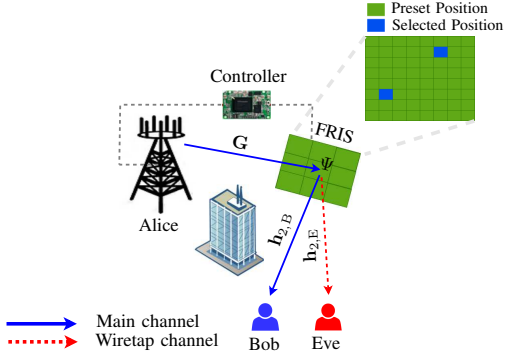}
\caption{System model for FRIS-enabled communication.
}
\label{SM}
\vspace{-4.5mm}
\end{figure}
We consider a \ac{FRIS}-assisted wiretap system with a transmitter Alice ($\mathrm{A}$) equipped with $L$ antennas, a single-antenna legitimate receiver Bob ($\mathrm{B}$), and a single-antenna eavesdropper Eve ($\mathrm{E}$). 
Direct line-of-sight links from $\mathrm{A}$ to $\mathrm{B}$ or $\mathrm{E}$ are blocked, so communication relies on the FRIS (see Fig.~\ref{SM}). The FRIS surface has area $A_T=N A_p$ in $\text{m}^2$, where $N=N_v N_h$ is the number of preset positions uniformly arranged in an $N_v \times N_h$ grid. 
Also, $A_p=d_{\rm H}d_{\rm V}$ denotes the area of each FRIS unit, with $d_{\rm H}$ and $d_{\rm V}$ being its horizontal width and vertical height, respectively, thus defining the physical footprint of each element. 
Each preset position is represented by $\mathbf{c}_n=[c_{x,n},c_{y,n}]^{\rm T}$, and the complete candidate set is $\mathcal{C}=\{\mathbf{c}_1,\dots,\mathbf{c}_N\}$. 
The surface is partitioned into $M$ disjoint subareas $\{\mathcal{S}_1,\dots,\mathcal{S}_M\}$, each assigned to one active fluid element that is restricted to move freely within its own subarea, but only among the $N/M$ discrete preset positions arranged as a local 2D grid. The position of the $m$-th element is $\mathbf{p}_m=[p_{x,m},p_{y,m}]^{\rm T}$, and the overall active configuration is $\mathcal{P}=[\mathbf{p}_1,\dots,\mathbf{p}_M]$, i.e., $M$ elements selected out of $N$. The received signal at $\mathrm{B}$ or $\mathrm{E}$ depends on the chosen configuration $\mathcal{P}$ and is expressed as
\begin{equation}
\label{eq1}
y_{i,(\mathcal{P}) }= \sqrt{P} \, \mathbf{h}_{2,i}^{\mathrm{H}} \bm{\Psi}_{(\mathcal{P})} \mathbf{G}_{(\mathcal{P})} \mathbf{w} x + n_i,
\end{equation}
where $i \in \{\mathrm{B},\mathrm{E}\}$ denotes either the legitimate or eavesdropper channels, $P$ is the transmit power at $\mathrm{A}$, $x$ the transmitted signal, and $n_i\sim \mathcal{C}\mathcal{N}(0,\sigma_i^2)$ the the additive white
Gaussian noise with variance $\sigma_i^2$. 
The maximum ratio transmission (MRT) BF vector at $\mathrm{A}$ is $\mathbf{w}$, while ${\bf{G}}_{(\mathcal{P}) }= [{\bf{g}_1,\dots,\bf{g}_{L}]} \in \mathbb{C}^{M\times L}$ and ${\bf{h}}_{2,i}=[ h_{2i,1},\dots,h_{2i,M}]^{\rm H} \in \mathbb{C}^{M\times 1}$ denote the $\mathrm{A}$-to-FRIS and FRIS-to-$i$ channels, respectively. 
The FRIS phase-shift matrix is $\bm{\Psi}_{(\mathcal{P}) }=\diag(e^{j\phi_{1}},\dots,e^{j\phi_{M}})$, assuming no reflection loss. Each element aligns its phase toward $\mathrm{B}$ as 
$\phi_m=-\angle(h_{2,\mathrm{B},m}^{\rm H})-\angle({\bf{g}}_m{\bf{w}})$, 
with $h_{2,\mathrm{B},m}^{\rm H}$ the $m$-th entry of ${\bf{h}}_{2,\mathrm{B}}^{\rm H}$ and ${\bf{g}}_m$ the $m$-th row of ${\bf{G}}_{(\mathcal{P})}$. 
The pair $(\mathbf{w},\bm{\phi}_m)$ is optimized to maximize the end-to-end SNR at $\mathrm{B}$. 
All FRIS links are modeled as Rayleigh fading with spatial correlation. Hence,
${\bf{h}}_{2,i}\sim \mathcal{C}\mathcal{N}\!\left(\bm{0}_{N},A_p\beta_{2,i}{\bf{R}}_{(\mathcal{P})}\right)$ and 
${\bf{g}}_{m}\sim \mathcal{C}\mathcal{N}\!\left(\bm{0}_{M},A\beta_1{\bf{R}}_{(\mathcal{P})}\right)$, 
for $m\in\{1,\dots,M\}$, where $\beta_1$ and $\beta_{2,i}$ denote the path-loss factors of the $\mathrm{A}$-to-FRIS and FRIS-to-$i$ links. 
The correlation matrix ${\bf{R}}_{(\mathcal{P})}\in\mathbb{C}^{M\times M}$ follows Jake’s model under rich scattering, so that its $(a,b)$-th entry is
\vspace{-1mm} 
\begin{equation}\label{eq4}
{\left [{\bf{R}}_{(\mathcal{P})}  \right ] }_{a,b}= J_0 \left ( 2\left \| {\bf{u}}_a-{\bf{u}}_b \right \| /\lambda \right )     \hspace{2mm}  a,b=1,\dots,M,
\vspace*{-1mm} 
\end{equation}
where $\lambda$ is the carrier wavelength, 
${\bf{u}}_\zeta = \left[ 0, {\rm mod}\!\left(\zeta - 1, M_h \right) d_{\rm H}, \; \left\lfloor (\zeta - 1)/M_v \right\rfloor d_{\rm V} \right]^{T}$ 
with $\zeta \in \{a,b\}$. 
Here, $M_h$ and $M_v$ represent the numbers of selected FRIS positions along the horizontal and vertical dimensions of the metasurface, respectively, so that $M = M_h \times M_v$. 
Based on these definitions and \eqref{eq1}, the received \ac{SNR} at both $\mathrm{B}$ and $\mathrm{E}$ can be written as
 \vspace{-2mm}
\begin{align}\label{eqsnr}
\gamma_{i,(\mathcal{P}) }=&\overline{\gamma}_i \,| \mathbf{h}_{2,i}^{\mathrm{H}} \bm{\Psi}_{(\mathcal{P})} \mathbf{G}_{(\mathcal{P})} \mathbf{w}|^2,
\end{align}
where, we define $\overline{\gamma}_i=P/\sigma_i^2$ as the average transmit SNR for $\mathrm{B}$ or $\mathrm{E}$. 
\vspace{-1mm}
\section{Problem Formulation and Algorithm Design}
\vspace{-0.5mm}
\subsection{Problem Formulation}
Here, the goal is to choose the most suitable \(M\) positions out of the total \(N\) preset candidates distributed over the FRIS surface so as to maximize the received \ac{SNR} at $\mathrm{B}$. 
Formally, let the complete set of $N$ candidate positions be 
$\mathcal{C} = \{\mathbf{c}_1, \mathbf{c}_2, \dots, \mathbf{c}_N\}$, 
from which we define a subset $\mathcal{P} \subset \mathcal{C}$ with cardinality $|\mathcal{P}| = M$, representing the positions of the $M$ active fluid elements. Accordingly, the optimization problem can be expressed as:
 \vspace{-1mm}
\begin{subequations}
\label{eq:opt_problem}
\begin{align}
(\rm{P1}): \quad & \max_{\mathcal{P} \subset \mathcal{C},\, |\mathcal{P}| = M} \quad \gamma_{i,(\mathcal{P})} \label{eq:opt_obj} \\
\text{s.t.} \quad
& \mathbf{p}_m \in \mathcal{S}_m, \quad \forall m = 1,\dots,M, \label{eq:pos_subarea} \\
& \|\mathbf{p}_m - \mathbf{p}_{m'}\|_2 \geq D, \quad \forall m \ne m', \label{eq:min_dist}
\end{align}
\end{subequations}
where $\mathbf{p}_m \in \mathbb{R}^2$ denotes the two-dimensional coordinate of the $m$-th selected fluid element, and the index $m'$ corresponds to another element different from $m$. 
Constraint~\eqref{eq:min_dist} ensures that a minimum separation distance $D$ is maintained between elements, thereby preventing electromagnetic coupling or physical overlap.
\vspace{-1mm}
\subsection{Optimization Algorithm Design}
\subsubsection{Q-Learning for Position Selection}
Due to the non-convexity of problem (P1), conventional methods may not reach the global optimum. 
Hence, we adopt a Q-learning scheme \cite{Qlearning}, where the FRIS controller acts as the agent, the states represent the spatial configuration of the fluid elements, and the actions correspond to selecting new positions from $\mathcal{C}$. 
The environment provides a reward defined as the end-to-end \ac{SNR} at $\mathrm{B}$, and the Q-table is updated iteratively. 
The update rule for a state-action pair $(s,a)$ at iteration $t$ is:
\vspace{-2mm}
\begin{align}
Q_{t+1}(s,a) = Q_t(s,a) + \alpha \Big[ r + \delta \max_{a'} Q_t(s',a') - Q_t(s,a) \Big],
\label{eq:q_update}
\end{align}
where $\alpha \in (0,1]$ is the learning rate, $\delta \in [0,1]$ is the discount factor, $r$ is the immediate reward, $s'$ is the next state, and $\max_{a'} Q_t(s',a')$ denotes the maximum Q-value attainable from the new state.  
Algorithm~\ref{alg:qlearning} outlines the proposed Q-learning-based procedure for optimizing the spatial configuration of the FRIS fluid elements. 
\begin{algorithm}[!t]
\scriptsize
\caption{Q-Learning-Based Position Optimization for FRIS}
\label{alg:qlearning}
\begin{algorithmic}[1]
\State \textbf{Input:} Total area $A_T = [X_{\min}, X_{\max}] \times [Y_{\min}, Y_{\max}]$, number of fluid elements $M$, minimum spacing $D$, power $P$, maximum episodes $E$, maximum steps per episode $T$, learning rate $\alpha$, discount factor $\delta$, exploration probability $\epsilon$.
\State \textbf{Initialization:} Initialize Q-table $Q(s,a)$ with zeros for all states $s$ and actions $a$. Partition the FRIS surface into $M$ subareas $\{\mathcal{S}_1, \dots, \mathcal{S}_M\}$ with spacing $\geq D$. Precompute channels: $\mathrm{A}$-to-FRIS $\mathbf{G} \in \mathbb{C}^{N \times L}$, FRIS-to-$\mathrm{B}$ $\mathbf{h}_{2,\mathrm{B}} \in \mathbb{C}^{N \times 1}$, and correlation matrix $\mathbf{R} \in \mathbb{C}^{N \times N}$.
\For {each episode $e = 1$ to $E$}
    \State Initialize state $s_0$ with random positions $\mathcal{P}^{(0)} = \{\mathbf{p}_1^{(0)}, \dots, \mathbf{p}_M^{(0)}\}$ within $\{\mathcal{S}_m\}$.
    \For {each step $t = 1$ to $T$}
        \State Select action $a_t$ (new position for one element) using an $\epsilon$-greedy policy from $Q(s_t,a)$.
        \State Apply action $a_t$ to obtain the new configuration $\mathcal{P}^{(t)}$ and the corresponding state $s_{t+1}$.
        \State Evaluate reward $r_t$ as the received SNR $\gamma_{\mathrm{B},(\mathcal{P}^{(t)})}$ based on \eqref{eqsnr}.
        \State Update Q-table using~\eqref{eq:q_update}. 
        \If {termination criterion is satisfied (e.g., convergence of $\mathcal{P}$ or maximum reward reached)}
            \State \textbf{break}
        \EndIf
    \EndFor
\EndFor
\State \textbf{Final greedy selection:} Starting from an empty configuration, sequentially select for each subarea $\mathcal{S}_m$ the action $a^\star = \arg\max_a Q(s,a)$, ensuring feasibility with respect to the spacing constraint $D$. The resulting configuration is denoted by $\mathcal{P}^\star$.
\State \textbf{Output:} Optimized configuration $\mathcal{P}^{\star} = \{\mathbf{p}_1^{\star}, \dots, \mathbf{p}_M^{\star} \}$ with maximum reward $\gamma_{\mathrm{B},(\mathcal{P}^{\star})}$ with final dimensions: $\mathbf{G} \in \mathbb{C}^{M \times L}$, $\mathbf{R} \in \mathbb{C}^{M \times M}$, $\mathbf{h}_{2,\mathrm{B}} \in \mathbb{C}^{M \times 1}$..
\end{algorithmic}
\end{algorithm}
\subsubsection{Distributed Phase Shift and Beamforming Optimization}
After obtaining the optimal set of positions $\mathcal{P}^\star$, we proceed to the joint optimization of the transmit BF vector at node $\mathrm{A}$ and the PS of the active FRIS elements. The corresponding problem is expressed as
\vspace{-1mm}
\begin{subequations}
\begin{align}
(\rm{P2}): \quad & \max_{\mathbf{w},\, \bm{\Psi}_{(\mathcal{P}^\star)}} \gamma_{\mathrm{B},(\mathcal{P}^{\star})} \label{eq:joint_opt_obj} \\
\text{s.t.} \quad & \|\mathbf{w}\|^2 \leq P_{\max}, \label{eq:joint_opt_c1} \\
& 0 \leq \phi_m < 2\pi,\quad \forall m = 1, \dots, M, \label{eq:joint_opt_c3}
\end{align}
\label{eq:joint_opt}
\end{subequations}
where $\gamma_{\mathrm{B},(\mathcal{P}^{\star})}$ denotes the end-to-end SNR at $\mathrm{B}$ given the fixed configuration $\mathcal{P}^\star$. To solve $(\rm{P2})$, we adopt a distributed alternating optimization method \cite{MRT}. In this approach, the transmit BF at $\mathrm{A}$ and the phase shifts at the FRIS are updated iteratively: in each iteration one variable is optimized while the other remains fixed, and the procedure continues until convergence or until a maximum number of iterations is attained. The proposed FRIS-specific algorithm is summarized in Algorithm~\ref{alg:distributed}. Finally, upon solving $(\rm{P1})$ and $(\rm{P2})$, the received in both $\mathrm{B}$ and $\mathrm{E}$ can be written as 
\begin{equation}
y_{i,(\mathcal{P}^\star)} = \sqrt{P}\,\mathbf{h}_{2,i}^{\mathrm{H}} 
\bm{\Psi}_{(\mathcal{P}^\star)}^\star 
\mathbf{G}(\mathcal{P}^\star)\,\mathbf{w}^\star x + n_i.
\label{eqoptima}
\end{equation}
Notice that $\bm{\Psi}_{(\mathcal{P}^\star)}^\star 
\mathbf{G}(\mathcal{P}^\star)\,\mathbf{w}^\star$ corresponds to the fully optimized end-to-end link towards $\mathrm{B}$. Consequently, the received SNR in both $\mathrm{B}$ and $\mathrm{E}$ nodes is given by
\begin{equation}
\label{SNRop}
\gamma_{i,(\mathcal{P}^\star)} 
= \overline{\gamma}_i\,
\left| \mathbf{h}_{2,i}^{\mathrm{H}}
\bm{\Psi}_{(\mathcal{P}^\star)}^\star
\mathbf{G}(\mathcal{P}^\star)\,
\mathbf{w}^\star \right|^{2}      
= \overline{\gamma}_i\,\left|h_i^\star \right|^{2},
\end{equation}
where $h_i^\star$ denotes the equivalent end-to-end FRIS channel after optimization.
\vspace{-2mm} 
\section{Performance Analysis}
\subsection{Approximate SNR Distributions}
Here, we employ the Maximum Likelihood Estimation (MLE) method to approximate the PDFs of the magnitude $\left| h_i^\star \right|$ with a single distribution, denoted as $H_i \sim \text{Nakagami}(\bm{\varphi}_{i}=\{\Omega_i,m_i\})$, where $\Omega_i$ denotes the average power and $m_i$ is the fading parameter. 
Let us then consider a training set vector ${\bf{h}}_i^\star=\{ h^\star_{i,j} \}_{j=1}^{t_{sp}}$, consisting of $t_{sp}$ independent samples of $h_i^\star$ obtained from~\eqref{SNRop}. The parameter vector $\bm{\varphi}_i$ is estimated by maximizing the log-likelihood function $L\!\left(H_i\mid\bm{\varphi}_i\right)$ associated with the PDF of $H_i$, based on the sample data set ${\bf{h}}_i^\star$. 
Accordingly, following~\cite{blume2002expectation}, the log-likelihood function is expressed as
\vspace{-2mm}
\begin{equation}\label{eq:eq8}
L\left(H_i \mid \bm{\varphi}_i\right) 
= \sum_{j=1}^{t_{sp}}\log\!\left[ f_{H}\!\left( h^\star_{i,j};\bm{\varphi}_i\right) \right].
\end{equation}
Then, the parameter estimates are obtained by setting the gradient of $L(H_i \mid \bm{\varphi}_i)$ with respect to each parameter to zero. 
Thus, the MLE estimator of an entry $\varphi_{i,t}\in\bm{\varphi}_i$, with $t \in \{1,2\}$, is given by
\begin{equation}\label{eq:eq9}
\frac{\partial L(H_i \mid \bm{\varphi}_i)}{\partial \varphi{i,t}}
= \sum_{j=1}^{t_{sp}} \frac{\partial}{\partial \varphi{i,t}} 
\log\!\left[f_{H_i}\!\left(h^\star_{i,j};\bm{\varphi}_i\right)\right] = 0.
\end{equation}
\begin{proposition}\label{Propos1}
Once the parameter vector $\bm{\varphi}_i$ is estimated with MLE, the PDF of $H_\mathrm{E}$ and the CDF of $H_\mathrm{B}$ of the received SNR can be approximated using a Gamma distribution via $\gamma_{i,(\mathcal{P}^\star)}=\overline\gamma_i|H_i|^2$.  So, the corresponding expressions are given by
\vspace{-1.5mm}
\begin{align}\label{eqEveBob}
f_{\gamma_{\mathrm{E},(\mathcal{P}^\star)}}(\gamma_E) &= \frac{\gamma_E^{\,k_E - 1}}{\Gamma(k_E)\, \overline\gamma_\mathrm{E}\,\theta_E^{k_E}} 
\exp\!\left(-\frac{\gamma_E}{\overline\gamma_\mathrm{E}\theta_E}\right), \\ \nonumber
F_{\gamma_{\mathrm{B},(\mathcal{P}^\star)}}(\gamma_\mathrm{B}) &= \frac{\Upsilon\!\left( k_\mathrm{B}, \frac{\gamma_\mathrm{B}}{\overline{\gamma}_\mathrm{B}\theta_\mathrm{B}} \right)}{\Gamma\!\left( k_\mathrm{B} \right)},
\end{align}
\end{proposition}
where $k_i=m_i$, and $\theta_i=\frac{\Omega_i}{m_i}$, with
\begin{align}\label{eq:parameters}
    \Omega_i = \frac{1}{t_{sp}}\sum_{j=1}^{t_{sp}} (h^\star_{i,j})^2 \hspace{4mm} \text{(a)}, \hspace{6mm} 
    m_i = \frac{1+\sqrt{1+\tfrac{4\Delta_i}{3}}}{4\Delta_i} \hspace{4mm} \text{(b)},
\end{align}
in which $\Delta_i$ is given by \eqref{eq:mMLNaka}.
\begin{proof}
    See Appendix~\ref{appendix1}.
\end{proof}
\begin{algorithm}[!t]
\scriptsize
\caption{Distributed Phase Shift and Beamforming Optimization}
\label{alg:distributed}
\begin{algorithmic}[1]
\State Initialize tolerance $\tau > 0$ and iteration index $k = 1$.
\State \textbf{BF initialization:} Estimate $\mathbf{G}_{(\mathcal{P}^\star)} \in \mathbb{C}^{M \times L}$ and set
$\mathbf{w}^{(k)} = \mathbf{1}_L/\sqrt{L} \in \mathbb{R}^{L \times 1}$.
\Repeat
    \State \textbf{PS update:} With $\mathbf{w}^{(k)}$ fixed, compute $ 
        \phi_m^{(k+1)} = -\angle({ h}_{2,\mathrm{B},m}^{\mathrm H})-\angle({\bf{g}}_m{\bf{w}}^{(k)}),
$ 
    and set $\bm{\Psi}_{(\mathcal{P}^\star)}^{(k+1)}= \mathrm{diag}\!\left( e^{j\phi_1^{(k+1)}}, \dots, e^{j\phi_M^{(k+1)}} \right)$.
    \State \textbf{BF update:} With $\bm{\Psi}_{(\mathcal{P}^\star)}^{(k+1)}$ fixed, update 
   $ 
        \mathbf{w}^{(k+1)} = \tfrac{\big( \mathbf{h}_{2,\mathrm{B}}^{\mathrm{H}} \bm{\Psi}_{(\mathcal{P}^\star)}^{(k+1)}\mathbf{G}_{(\mathcal{P}^\star)}  \big)^{\mathrm H}}{\norm{ \mathbf{h}_{2,\mathrm{B}}^{\mathrm{H}} \bm{\Psi}_{(\mathcal{P}^\star)}^{(k+1)} \mathbf{G}_{(\mathcal{P}^\star)}  } } .
    $ 
    \State $k \gets k + 1$.
\Until{fractional increase in $\gamma_{\mathrm{B}}(\mathcal{P}^\star)$ $< \tau$ or maximum iterations reached}
\State \textbf{Output:} Optimized $\mathbf{w}^\star$ and $\bm{\Psi}^\star_{(\mathcal{P}^\star)}$.
\end{algorithmic}
\end{algorithm}
\vspace{-2mm}
\subsection{Secrecy Outage Probability}
We consider a passive wiretap scenario where the eavesdropper's CSI is unavailable at Alice. Hence, Alice transmits with a fixed secrecy rate $R_{\mathrm{S}}$. The secrecy capacity is defined as $ 
C_\mathrm{S}=\max\{C_\mathrm{B}-C_\mathrm{E},0\}$,
where $C_\mathrm{B}=\log_2(1+\gamma_\mathrm{B})$ and $C_\mathrm{E}=\log_2(1+\gamma_\mathrm{E})$ are the channel capacities at $\mathrm{B}$ and $\mathrm{E}$. Secure transmission holds if $R_\mathrm{S}\leq C_\mathrm{S}$; otherwise, a secrecy outage occurs. Thus, the secrecy outage probability (SOP) is $\text{SOP}=\Pr\{C_\mathrm{S}<R_{\mathrm{S}}\}$, and a tight lower bound is given by~\cite{Osorio2023_PLS}
\begin{align}\label{eq15}
\text{SOP}_{\text{L}}
 =\int_{0}^{\infty} F_{\gamma_{\mathrm{B},(\mathcal{P}^\star)}}\left(2^{R_{\mathrm{S}}}\gamma_\mathrm{E}\right) 
f_{\gamma_{\mathrm{E},(\mathcal{P}^\star)}}(\gamma_E)d\gamma_\mathrm{E}.
\end{align}
\begin{proposition}\label{Propo2}
For FRIS-assisted MISO systems, the $\text{SOP}_{\text{L}}$ can be approximated by
\begin{subequations}\label{eq17}
\begin{align}
\text{SOP}_{\text{L}} \approx \rho^{k_\mathrm{B}}
\frac{2^{k_\mathrm{B}R_{\mathrm{S}}}}{k_\mathrm{B}\,\mathcal{B}(k_\mathrm{B},1)}
\, {}_2F_1\!\left(k_\mathrm{B}+1,k_\mathrm{B};1+k_\mathrm{B};-2^{R_{\mathrm{S}}}\rho\right),
\end{align}
\end{subequations}
\end{proposition}
where, $
\rho \triangleq \frac{\overline{\gamma}_\mathrm{E}\theta_\mathrm{E}}{\overline{\gamma}_\mathrm{B}\theta_\mathrm{B}}
$, $\mathcal{B}\left (\cdot,\cdot\right )$ is the beta function, and ${}_2F_1\left(\cdot,\cdot;\cdot;\cdot\right)$ denotes the Gauss hypergeometric function.
\begin{proof}
The result is obtained from~\cite[Eq.~(7)]{Kong} after appropriate substitutions and mathematical manipulations.
\end{proof}

\section{Numerical results and discussions} \label{sect:numericals}
This section shows numerical results to validate the analytical framework and assess the secrecy performance of FRIS against conventional RIS schemes. Specifically, two FRIS modes are examined: 
$1)$ FRIS with only selected positions optimized (SPO), and $2)$ FRIS with joint optimization of element positions, transmit BF, and PS (SPO+BF+PS). 
For comparison purposes, the FRIS is evaluated against two RIS baseline configuration: 
$1)$ a \emph{Conventional RIS}, with $M$ fixed elements positioned at the subregion centers of the FRIS grid, thereby spanning the full aperture area., and $2)$ a \emph{Compact RIS}, where $M$ elements form a dense rectangular array with spacing $\leq \lambda/2$, resulting in a smaller aperture but stronger spatial correlation. For a fair comparison, both RIS baselines employ the same number of elements $M$ as the FRIS and are tested with either random PS or optimized BF and PS. Unless specified otherwise, the FRIS parameters are: carrier frequency $f_c=2.4$~GHz ($\lambda=0.125$~m), correlation matrix with $d_{\rm H}=d_{\rm V}=\lambda/3$, aperture size $A_T=2\,\text{m}\times2\,\text{m}$ ($N_{\rm v}=N_{\rm h}=48$), path-loss exponents $\beta_1=\beta_i=-40$~dB, secrecy rate threshold $R_\mathrm{S}=1$~bps/Hz, and $\overline{\gamma}_\mathrm{E}=12$ dB. The Q-learning algorithm adopts standard parameters: learning rate $\alpha=0.1$, discount factor $\delta=0.9$, exploration probability $\epsilon=0.1$, and $E=400$ episodes. For the MLE procedure, $t_{\rm sp} = 10^4$ realizations are used to form the training set in all cases. Monte Carlo simulations over channel realizations are conducted to validate the analysis, while Moment Matching (MoM) is included as a benchmark for SOP approximation.
\begin{figure*}[ht!]
\vspace{-5.5mm}
\centering
\psfrag{A}[Bc][Bc][0.5]{$\rho=5$ dB}
\psfrag{B}[Bc][Bc][0.5]{$\rho=-5$ dB}
\psfrag{C}[Bc][Bc][0.5]{no EMI}
\psfrag{D}[Bc][Bc][0.5]{$\rho=5$ dB}
\subfigure[FRIS spatial configuration with $M = 16$ and $L=2$.]{\includegraphics[width=0.32\textwidth, height=0.18\textheight]{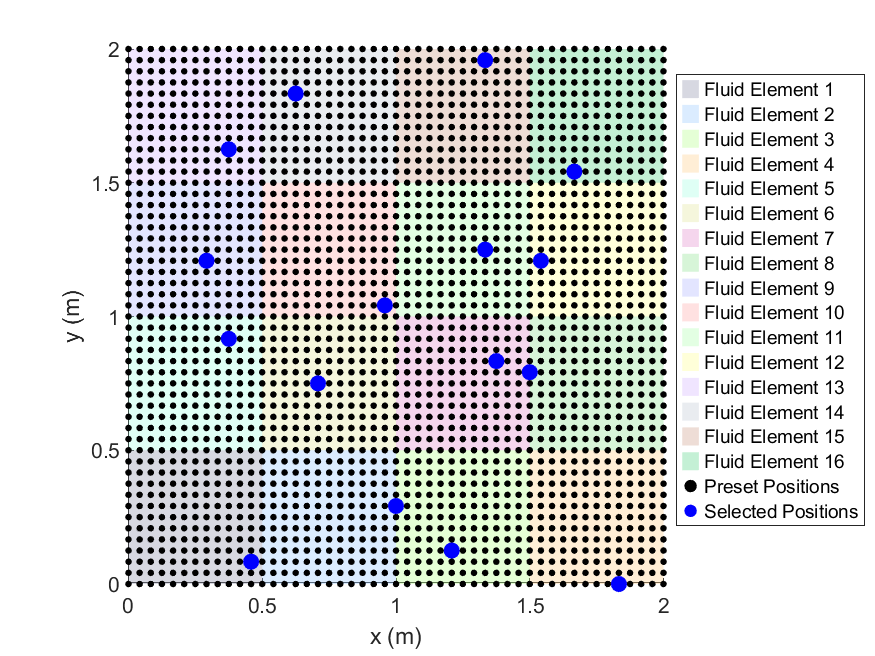}} 
\psfrag{E}[Bc][Bc][0.5]{$\rho=20$ dB}
\psfrag{A}[Bc][Bc][0.5]{$\rho=20$ dB}
\psfrag{B}[Bc][Bc][0.5]{$\mathrm{\textit{N}=324, 144}$}
\psfrag{C}[Bc][Bc][0.5]{$\sigma_{\text{EMI},\mathrm{B}}^2=-45$ dBm}
\psfrag{D}[Bc][Bc][0.5]{$\mathrm{\textit{N}=144, 324}$}
\subfigure[OP vs. $\overline{\gamma}_\mathrm{B}$ for $M=36$ fluid elements and $L \in \{2,6\}$. ]{\includegraphics[width=0.32\textwidth]{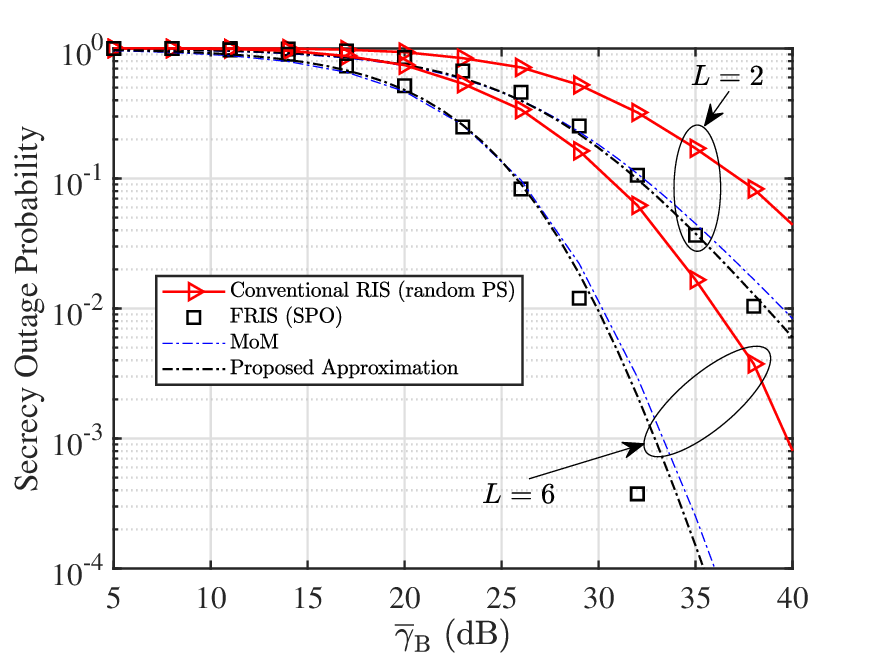}}
\psfrag{A}[Bc][Bc][0.5]{$\mathrm{\lambda/2}$}
\psfrag{B}[Bc][Bc][0.5]{$\mathrm{\lambda/5}$}
\psfrag{C}[Bc][Bc][0.5]{$\mathrm{\lambda/3}$}
\subfigure[SOP vs. $\overline{\gamma}_\mathrm{B}$ for $M=196$ and $L=4$.]{\includegraphics[width=0.32\textwidth]{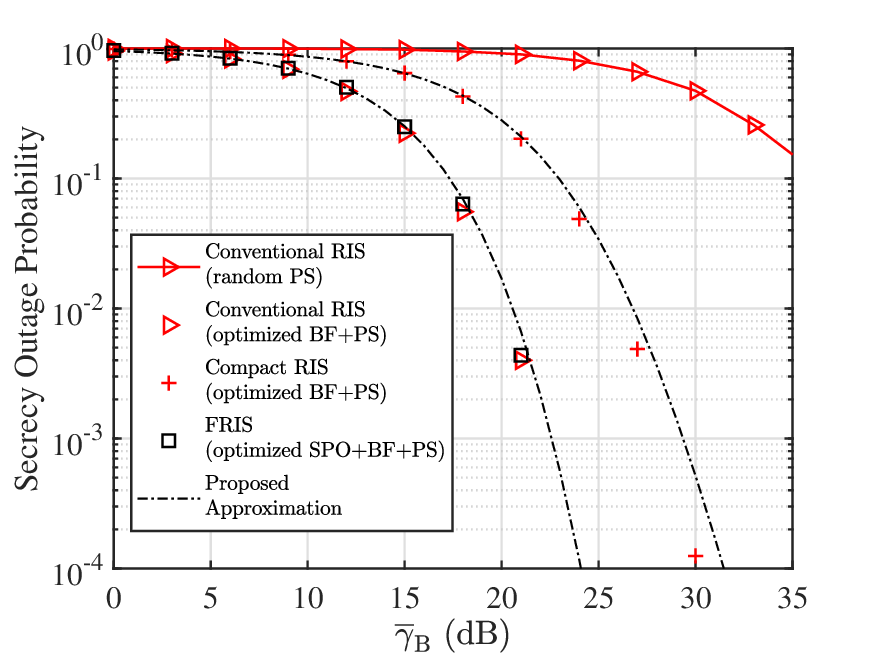}}
    \caption{Spatial configuration and OP achieved for FRIS-aided Multiple-Input Single-Output (MISO). In all figures, the dashed lines represent analytical solutions, while markers correspond to MC simulations. Also, for a fair comparison, the RIS is configured with the same number of elements $M$ as the FRIS in all cases. }
    \label{figen}
    \vspace{-4mm}
\end{figure*}
\begin{figure}[!t]
\vspace{-2.5mm}
\centering
\psfrag{A}[Bc][Bc][0.5]{$\rho= 30, 25, 20$ dB}
\psfrag{B}[Bc][Bc][0.5]{$\rho=40$ dB}
 \includegraphics[width=60mm]{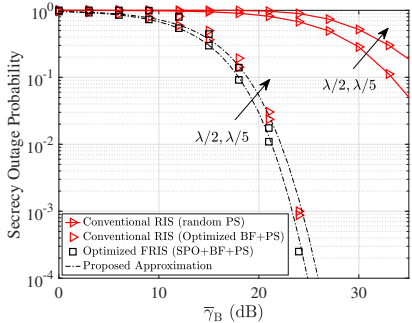}
\caption{SOP vs. $\overline{\gamma}_\mathrm{B}$ for $L=4$, $M=100$, and different inter-element spacings between fluid elements. Markers denote MC simulations whereas the solid lines represent analytical solutions.}
\label{fig5f}
\vspace{-4.5mm}
\end{figure}

Figure~\ref{figen}a illustrates the spatial configuration of the FRIS 
with $M=16$ fluid elements deployed within a square region. The background grid corresponds to the $N$ uniformly distributed candidate positions, ensuring fine spatial granularity across the surface. The blue markers highlight the active elements selected by the proposed secrecy-aware optimization framework, while the inactive candidate points are shown in black. From a PLS perspective, this adaptive positioning introduces spatial randomness that increases the eavesdropper’s channel uncertainty, effectively reducing the SOP compared to conventional and compact RIS architectures. However, the performance gains become less pronounced when 
contrasted with fully optimized RIS designs, where BF and PS 
strategies are employed.

In Fig.~\ref{figen}b, we compare the SOP performance of FRIS with spatial position optimization (SPO) against a conventional RIS with random PS, considering $M=36$ fluid elements and $L \in \{2,6\}$ transmit antennas. In this setup, neither FRIS nor RIS benefit from BF or PS optimization, thereby isolating the impact of the spatial geometry on the secrecy performance. From a PLS 
point of view, it is evident that FRIS with optimized positions provides a significant reduction in SOP compared to conventional RIS, especially in the high-SNR regime. This gain comes from the ability of FRIS to introduce spatial 
randomness, which increases the uncertainty of the eavesdropper’s channel. Moreover, as $L$ increases from $2$ to $6$, the secrecy performance improves notably, confirming that multiple transmit antennas and spatial diversity 
further enhance PLS. The figure also includes analytical benchmarks such as the Method of Moments (MoM) and the proposed approximation, both of which closely track the simulated SOP, with the proposed approximation exhibiting the 
best accuracy. These results validate the effectiveness of the proposed model and highlight the advantage of FRIS-assisted systems over conventional RIS in 
terms of secrecy performance.

In Fig.~\ref{figen}c, we evaluate the SOP 
performance of FRIS under full optimization, i.e., combined SPO, BF, and PS design, with 
$M=196$ fluid elements and $L=4$ BS antennas. The results are contrasted against multiple benchmarks: a Conventional RIS with random PS, an optimized Conventional RIS (BF+PS), and a Compact RIS (BF+PS). From the curves, it is evident that the optimized FRIS achieves a clear SOP advantage over both the 
Conventional RIS and the Compact RIS. This gain is particularly notable in the 
moderate-to-high SNR regime, where the spatial flexibility of FRIS mitigates 
the correlation losses that limit compact deployments. In contrast, when 
comparing FRIS with an optimized Conventional RIS (BF+PS), the performance gap 
becomes marginal, suggesting that once BF and PS are carefully designed, the 
additional spatial optimization introduced by FRIS provides only limited benefits in terms of SOP. From a PLS standpoint, these findings confirm that the main strength of FRIS lies in scenarios where PS optimization is restricted or impractical, as the spatial adaptation of fluid elements introduces channel randomness that increases the eavesdropper’s uncertainty. Meanwhile, the 
consistent superiority of FRIS over the Compact RIS highlights the importance 
of spatial diversity and reduced correlation in secrecy-oriented designs.

In Figure~\ref{fig5f}, we evaluate the SOP 
performance for both FRIS and conventional RIS under full optimization, 
considering $L=4$, $M=100$, and different inter-element spacings between fluid 
elements. Here, it is observed that by reducing the spacing (e.g., $\lambda/5$) 
increases the spatial correlation among reflecting elements, which in turn 
reduces the channel diversity and negatively impacts secrecy performance. This 
effect is clearly visible in the curves, where smaller spacings result in 
higher SOP values compared to the reference case of $\lambda/2$. Nevertheless, 
the optimized FRIS (SPO+BF+PS) consistently outperforms the optimized 
conventional RIS (BF+PS), highlighting that the spatial adaptability of FRIS 
can mitigate part of the secrecy degradation caused by reduced inter-element 
distances. In contrast, the conventional RIS with random PS yields the poorest 
performance due to the absence of any optimization. It is also worth noting that, when both FRIS and conventional RIS employ 
optimized BF and PS, the additional spatial configuration flexibility of FRIS 
does not translate into a substantial SOP gain. This observation is consistent 
with earlier results, confirming that the secrecy benefit of FRIS becomes less 
pronounced when RIS is already equipped with full optimization. Hence, the main 
advantage of FRIS lies in scenarios where PS optimization is restricted or 
impractical, since in such cases spatial adaptability introduces channel 
randomness that effectively enhances secrecy performance.

\vspace{-1mm}
\section{Conclusions}
In this work, the SOP performance of FRIS had been investigated and compared 
with both conventional and compact RIS architectures. It had been shown that 
the additional degree of freedom introduced by spatial position optimization 
had allowed FRIS to consistently outperform conventional RIS with random phase 
shifts, as well as compact RIS, which had suffered from higher spatial 
correlation and reduced diversity. Furthermore, results had indicated that FRIS 
could partially mitigate the secrecy degradation caused by reduced inter-element 
spacing, highlighting its robustness under practical deployment constraints. 
However, when conventional RIS had been equipped with optimized BF and 
Ps strategies, the spatial adaptability of FRIS had not provided 
significant additional SOP gains. Overall, FRIS had demonstrated clear 
advantages in scenarios where phase optimization was restricted or impractical, 
thus reinforcing its potential as a promising architecture for enhancing 
physical layer security.

\appendices
\section{Proof of Proposition \ref{Propos1}}
\label{appendix1} 
Here, we provide a derivation of the estimation of the distribution parameters of the Nakagami-$m$ model that approximates the PDF of $H_i$. By considering a training set vector ${\bf{h}}_i^\star=\{ h^\star_{i,j} \}_{j=1}^{t_{sp}}$ obtained from~\eqref{SNRop}, 
we approximate the PDF of $H_i$ by a Nakagami-$m$ distribution, expressed as
\begin{equation}\label{eq:Nakagami}
f_{H_i}(h;\bm{\varphi}_i)=\frac{2 m_i^{m_i} h^{2m_i-1}}{\Gamma(m_i)\,\Omega_i^{m_i}}
\exp\!\left(-\frac{m_i h^2}{\Omega_i}\right),
\end{equation}
where the parameter vector is $\bm{\varphi}_i=(m_i,\Omega_i)$. 
Now, to estimate $\Omega_i$, we substitute \eqref{eq:Nakagami} into the log-likelihood derivative \eqref{eq:eq9} for the parameter $\varphi_{i,t}=\Omega_i$, which directly leads to the result in~\eqref{eq:parameters}a. Similarly, solving \eqref{eq:eq9} with the PDF in \eqref{eq:Nakagami} for the parameter $\varphi_{i,t}=m_i$ yields
\begin{equation}\label{eq:mMLNaka}
\log(m_i)-\psi(m_i)= 
\underbrace{\log\!\left[\frac{1}{t_{sp}}\sum_{j=1}^{t_{sp}} (h^\star_{i,j})^2 \right]
-\frac{1}{t_{sp}}\sum_{j=1}^{t_{sp}}\log \left( (h^\star_{i,j})^2 \right)}_{\Delta_i},
\end{equation}
where $\psi(\cdot)$ is the digamma function~\cite[Eq. (6.3.1)]{abramowitz1972handbook}. Since \eqref{eq:mMLNaka} admits no closed-form solution for $m_i$, we approximate $\psi(\cdot)$ using its asymptotic expansion~\cite[Eq. (6.3.18)]{abramowitz1972handbook}. 
By applying a second-order expansion and performing some algebraic manipulations, the closed-form approximation of $m_i$ is obtained as given in~\eqref{eq:parameters}b. This completes the proof.





\bibliographystyle{ieeetr}
\bibliography{bibfile}

\end{document}